*Original Article*

# Comparative Analysis of AWS Model Deployment Services

Rahul Bagai

*Senior Software Engineer, AssemblyAI, Inc.*

*Corresponding Author : rahulbagai@ieee.org*

***Abstract*** *- Amazon Web Services (AWS) offers three important Model Deployment Services for model developers: SageMaker, Lambda, and Elastic Container Service (ECS). These services have critical advantages and disadvantages, influencing model developers' adoption decisions. This comparative analysis reviews the merits and drawbacks of these services. This analysis found that Lambda AWS service leads in efficiency, autoscaling aspects, and integration during model development. However, ECS  was found to be outstanding in terms of flexibility, scalability, and infrastructure control; conversely, ECS is better suited when it comes to managing complex container environments during model development, as well as addressing budget concerns- it is, therefore, the preferred option for model developers whose objective is to achieve complete freedom and framework flexibility with horizontal scaling. ECS is better suited to ensuring performance requirements align with project goals and constraints. The AWS service selection process considered factors that include but are not limited to load balance and cost-effectiveness. ECS is a better choice when model development begins from the abstract. It offers unique benefits, such as the ability to scale horizontally and vertically, making it the best preferable tool for model deployment.*

## 1. Introduction

In the rapidly evolving domain of Machine Learning (ML), the efficiency and effectiveness of model deployment services are pivotal.

Amazon Web Services (AWS) offers three robust solutions—SageMaker, Lambda, and Elastic Container Service (ECS)—each tailored to different aspects of deployment scalability, cost-efficiency, and integration capabilities [30].

This comparative analysis delves deep into these services, highlighting their unique advantages and potential drawbacks to help developers and organizations make informed decisions. This analysis reveals that while Lambda excels in efficiency and integration for simpler, event-driven applications, ECS offers unmatched flexibility and control for complex environments. Meanwhile, SageMaker provides a comprehensive, managed experience ideal for large-scale ML deployments [29].

By bridging the information gap with a thorough examination of these tools, this study aims to equip with the knowledge to choose the most suitable AWS service for specific model deployment needs, enhancing performance and cost-effectiveness.

## 2. AWS SageMaker

Amazon SageMaker is a completely managed Amazon Web Service (AWS) service that enables scientists, developers, and practitioners to quickly build, train, and deploy machine learning solutions on the cloud. It contains a range of functionalities and elements to assist with the ML process [2]. In addition, it offers a Jupyter Notebook environment through its hosted services to support data exploration and model development based on existing machine learning algorithms and frameworks that reduce the time needed for model development [2]. Besides, it will simplify the training process by caring for the base infrastructure and altering the models for best performance. At the end of the modeling exercise, the SageMaker models can be deployed directly onto the production system through auto-scaling and A/B testing that are built into it. Thus, AWS SageMaker enables the building of machine-learning apps that can be applied to different model development situation requirements [11].

Figure 1 shows how Amazon SageMaker comprises different components in creating machine learning models, starting from the strings and moving to the deployment stage. To mention a few examples, SageMaker Studio is a collaborative environment for developing models, while SageMaker Pipelines assists in creating ML workflows.

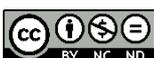

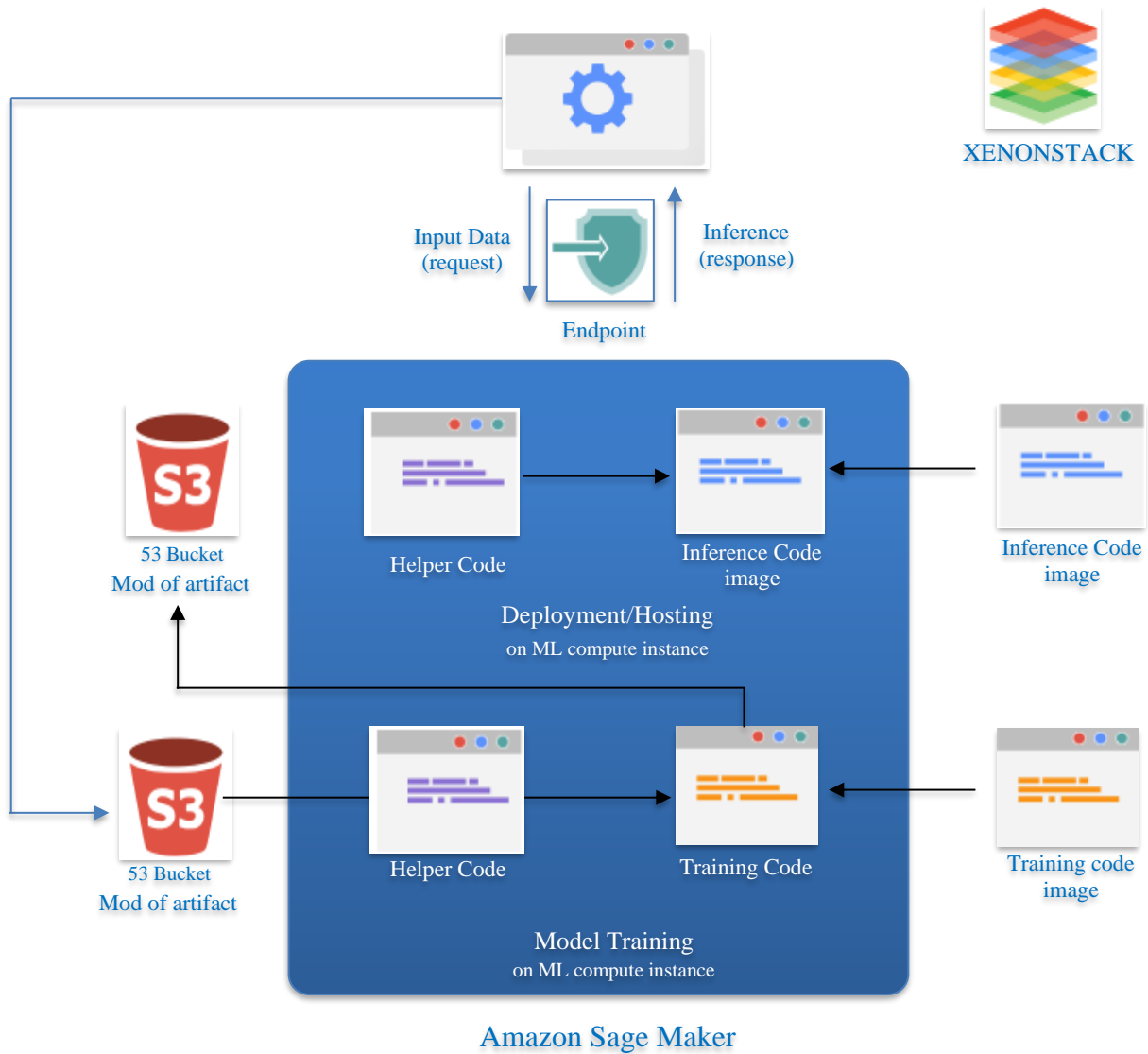

**Fig. 1 Amazon SageMaker architecture [35]**

In the same light, the Scalable Training Process of IT, using models and scaling them at the time of deployment, is all now possible. Model monitoring and management are among the most critical components of a well-used product built on the SageMaker ecosystem, as these components determine the success in both the delivery and governance of Machine Learning experiments.

From a practical aspect, interoperability, together with other possible AWS services, simplifies the process of use and scalability (S3). For instance, one can open the data on Amazon S3 directly during model training and training without any hurdles and utilize the AWS Glue application when running your data sources' ETL (Extract, Transform, and Load) tasks in cloud-based platforms such as Redshift [7]. Integration accomplished between SageMaker into AWS Lambda and API Gateway also makes the deployment of RESTful APIs. This shows that SageMaker has a wide range of integration with different machine learning models, which could be efficiently scaled across varying industry segments [4]. Ultimately, Amazon SageMaker provides a robust and complete foundation for deploying ML models on an infrastructure and services provided by AWS.

### 2.1. Strengths
#### 2.1.1. Scalability in AWS SageMaker
Amazon SageMaker is scalable and can handle cases of large-scale deployments easily. SageMaker comes bundled with an expanded Machine Learning Operations (MLOps) model for administering Big Data and ML algorithms at the enterprise level [16]. This scalability becomes essential in cases like energy supply when dealing with large volumes of raw data and when implementing sophisticated machine learning models, which are the norm.

Using the AWS cloud service, SageMaker simplifies the scaling process by preventing disruptions that happen in the process and allows organizations to handle them without any performance impact. Leveraging its high computational efficiency, it built a recommendation system for a mobile application, where AWS Lambda is the core [6].
#### 2.1.2. Infrastructure and Ease of Use

One of the other advantages of Amazon Web Services' SageMaker is its managed infrastructure, which heavily decreases operational hassles. SageMaker is a single platform that provides end-to-end ML pipelines, including data collection and model deployment [19]. This managed platform offers an easy way to provision and handle resources, allowing the data experts and developers to devote more time to model building and deployment tasks than infrastructure management. Besides modularity, SageMaker's extensibility and compatibility with pre-established workflows [17].

He ported his old algorithms to SageMaker's platform for better efficiency and functionality. This user-friendliness leads to faster development and reduces operational costs.

*2.1.3. Built-In Algorithms and Frameworks*
The more developed a feature that the Amazon Web Services SageMaker has, the more it contains a wide range of built-in ML techniques and frameworks that can simplify many projects. The usefulness of [19] was demonstrated in time series forecasting, AutoML, and anomaly detection so that the SageMaker pre-built tools and models, including optimized and premade algorithms and frameworks.

This is reflected in Soncin's work, where the ML models were easily migrated due to the pre-existing algorithms from SageMaker, which aided the migration to a more modern, efficient environment [16]. Consequently, the built-in features of SageMaker speed up the development process, thus making it more suitable for deploying AI models on AWS by accelerating the experiments.

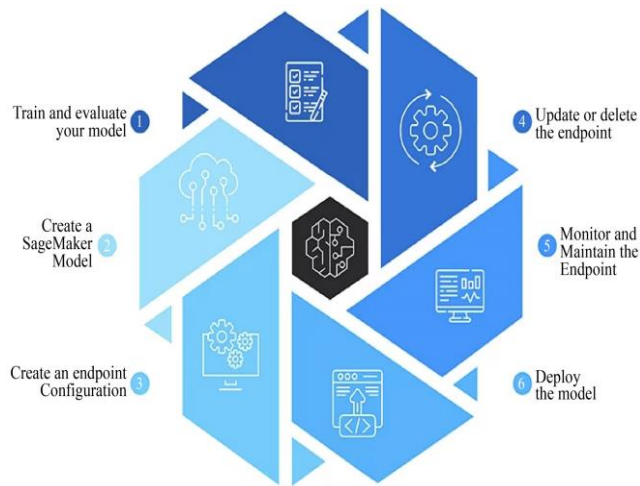

**Fig. 2 AWS SageMaker [32]**

Figure 2 highlights how the machine learning models are deployed on AWS SageMaker. Demonstrating SageMaker's ability to accommodate all the other AWS services and its managed infrastructure strengthens the system's ease of use. It shows how the machine learning end-to-end work cycle has been rebranded by SageMaker, facilitating data scientists and developers to simplify deployment activities.

*2.2. Weaknesses*
*2.2.1. Costing of AWS SageMaker*
Amazon SageMaker is an ML platform encompassing tools such as model training, deployment, and management. These are all in one place [25]. However, one disadvantage of this machine-learning device is the price. SageMaker can also become expensive. Regarding big projects or deployments, computational power is too high [9]. Pricing is contingent upon various factors, such as training instances, inference endpoints, and storage resources, which may account for high variation in project costs [11]. In addition, its strong skills and cost-effectiveness must be evaluated from the point of view of those companies that face financial difficulties or budget-oriented tenders.

*2.2.2 AWS SageMaker Offers Limited Customization*
Regarding SageMaker by AWS, customization is limited to the flexibility feature for advanced users. While not all algorithms and frameworks for typical ML tasks are built-in, one risk of using the platform is the limitations posed by predefined configurations that SageMaker provides [6]. Customization needs and deploying beyond the capabilities of the SageMaker-managed infrastructure may require developing non-standard and rather complex solutions [9]. However, basic templates may not be sufficient, especially if some projects need a custom setup that goes beyond the functionality of AWS.

*2.2.3. Performance Considerations on AWS SageMaker*
Model deployment can be divided into the speed of deploying the model, the speed at which the model can perform inference, and the overall performance of the model. However, when considering the model's responsiveness, it is important to monitor both latency and throughput to meet performance needs carefully. The narrow optimization of models to support real-time inference and applications with hard latency limits can be needed [11]. Besides, the kind of inferences and configurations in SageMaker determine the speed and resource consumption, respectively [8]. As a result, this implies that performance benchmarking and evaluation should be done for any SageMaker deployment issues related to bottlenecks or latency issues.

*2.2.4. Cost Analysis of Amazon AWS SageMaker*
Conducting economic feasibility is inevitable when expecting the return on investment of applying machine learning models through AWS SageMaker. The pricing structure associated with SageMaker depends on different components, such as the instance types, duration of the sessions, and the storage price that determines the price models [6]. Likewise, organizations, in doing so, should consider the operational costs in planning their deployment strategies on AWS, which also involve the Total Cost of Ownership (TCO), including operational overhead expenses from operations, while also deciding on the model implementation decisions being made informally [24]. Thus, tighter control of investments will allow organizations to do detailed cost analyses, which will put resources where they are wanted and develop Return On Investment (ROI) from machine learning.

## 3. AWS Lambda
### 3.1 AWS Lambda overview and use cases
AWS Lambda is a serverless computing service from Amazon Web Services – it runs the code without having to provision or manage servers. For example, Choudhary mentioned that the lambda function could be triggered via HTTP requests. It could modify data in Amazon's S3 buckets or update tables in the DynamoDB database. Hence, it could be considered that Lambda is a relevant option for machine learning model deployment using an event-driven architecture. A case in point may be to use Lambda functions to convert data in S3 buckets, respond to HTTP requests for inference models, and stream data in real-time [1]. One of the crucial advantages of this system is that machine learning applications can be developed to be scaled quickly and cost-effectively without the pressure of infrastructure management.

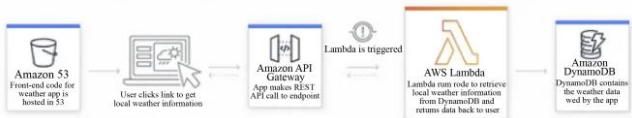

**Fig. 3 AWS lambda workflow** [36]

Figure 3 shows the AWS Lambda workflow, which begins with the events that occur, followed by the execution of the code, and ends with the results produced. On the other hand, the events can be triggered from different sources like the S3, API Gateway, DynamoDB, or IoT devices, which will further the work of assigned jobs to the Lambda functions.

### 3.2. AWS Lambda's Features and Benefits
AWS Lambda is one of the best options for a machine learning model because of its many cool features and advantages. This matches the needs of apps that use automatic scaling and availability features, avoiding breakages in consistency due to traffic spikes [1]. Besides, it adopts a serverless architecture model, eliminating the operational challenges besides the difficulties developers face when setting up servers [1]. Furthermore, it would help if one took note of the pay-per-use plan, where customers are charged only when they use their computing time, thus making it cost-effective for both small and large deployments.

Apart from that, Lambda is integrated with multiple services such as API Gateway, DynamoDB, and S3; therefore, developers can build very resource-efficient serverless frameworks and integrate them with any other cloud components of choice [1]. Highly scalable and flexible, the AWS Lambda environment allows developers to use various programming languages and secure identity management and development and deployment controls. Hence, it becomes a decent platform for deploying machine learning models.

### 3.3. Strengths
#### 3.3.1. High Efficiency of AWS Lambda
Using AWS Lambda virtual protocols, one can execute the codes on a server without thinking of profitability being affected by the allocation or operation of servers. These capabilities are supported by a pay-as-you-use pricing model suitable for rare peaks or poor traffic conditions [13]. This will help optimize the costs as it averts the need for idle and maintenance expenses by matching the consumption to expenditure. Switching to server-free models where cloud infrastructure is dynamically scaled as workload fluctuates, corporations can attain remarkable cost savings through Lambda [10]. Hence, it is feasible for various use cases, including small-scale applications and large-scale corporate solutions.

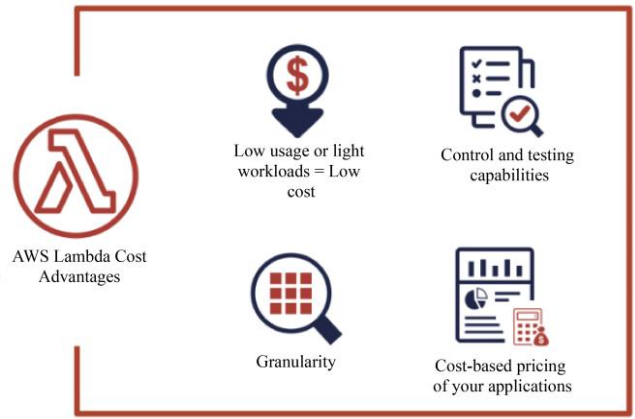

**Fig. 4 Lambda pricing** [33]

Figure 4 illustrates how AWS Lambda has pay-per-use pricing, in which businesses will be charged for the actual time their functions run. This reduces costs since these models do not need investing in additional provisioning.

#### 3.3.2. Auto-scaling Capability of AWS Lambda
AWS Lambda's strengths lie in its auto-scaling capabilities. Lambda automatically scales to support the maximum number of requests, and no one is required to be involved in the process [8]. With this automatic scaling feature, organizations can easily cope with higher or lower capacities without capacity planning and provisioning. When resource demand increases dynamically at peak times, Lambda will assign additional resources to handle incoming requests to maintain performance with minimal delays [26]. On the contrary, when they do not have any operations, they will allow the reduction of Lambda automatically to ensure that their expenses are minimized while resource usage is maintained. This feature makes it a convenient choice over others in unsteady state workloads such as those in apps where scalability and agility are primarily required.

#### 3.3.3. Integration with AWS Ecosystem
AWS Lambda can be integrated within the broader stream of AWS-related services for which only a few services in the ecosystem may be necessary for designing and deploying serverless applications. For example,

Lambda functions can be considered through Amazon S3, Amazon DynamoDB, and Amazon API Gateway, which facilitate processing, storing, and communication [10]. Surprisingly, this intricate integration attains the goal of simple serverless architectures where joint work is executed among all the AWS services. Moreover, it has an integration feature with Amazon monitoring and logging facilities, permitting access to essential function performances and operational metrics. By adopting AWS services, organizations could gain a foothold in developing faster applications, more extensive scalability, and more agility in deploying serverless applications.

### 3.4. Weaknesses
*3,4.1. Issues of Time Constraints and Performance Restrictions*

Amazon's Lambda AWS enables the utilization of serverless technologies and is, on the one hand, a challenging factor in employing complex machine learning implementations or data processing jobs due to set time outs. Lambda function will have an upper ceiling of their run-time, up to 15 minutes [5]. This could involve splitting the tasks into smaller units or finding other means to run the programs that will not prove successful because of the brief duration. Furthermore, these resource constraints (such as memory, CPU, and storage) imposed by AWS Lambda negatively affect the performance of resource-intensive jobs [5]. These constraints should be factored in when designing and deploying applications.

*3.4.2. AWS Lambda's Performance Anxiety*

AWS Lambda should not be underestimated in performance evaluation, such as introducing machine learning models and other computational tasks. Lambda function latency comprehension is required because it affects the system-level performance [12]. Lambda is good in parallel scaling, but cold start latency and network overhead can cause a real-time application to be unresponsive. As such, benchmarking can help improve the performance of lambda functions by considering factors such as invocation rate, payload size, concurrency settings, and depending on a specific use case [27]. In addition, just monitoring the lambda functions to ensure they run at optimum levels would make all metrics perform with less waste of resources.

*3.4.3. Costing for AWS Lambda*

A complete cost analysis is one of the essential parts of the cost analysis that should be done before deciding on the economies of deploying machine learning models or data pipelines on the AWS Lambda service. [8] noted that the cost factors for the operations count, memory allocations and function call duration should be part of the decision-making process. The pay-per-use principle is favorable for both the infrequent and dormant Lamdas, but these users may spend much more when they are on or very busy [25]. Hence, predicting how Lambda and its correlative technologies, like serverless computing, will be employed is fundamental for companies to maximize expected business value and minimize costs. Some price comparisons between Lambda's "serverless compute" pricing models and the deployment alternatives such as SageMaker and ECS can show that Lambda is more cost-effective for some tasks.

## 4. AWS Elastic Container Service (ECS)
### 4.1. Use of AWS Elastic Containers Service (ECS) and its Constituents

If a user is looking for options for managing models in ML containerized applications, Amazon Elastic Container Service (ECS) will be a good choice for its flexibility. Flexibility, scalability, and integration with other AWS services are the main advantages of instances. Another attractive feature of ECS is that it allows companies to place modeling inference logic inside reusable containers across different environments [14]. Thus, the use-case of ECS is aiding organizations in packing, deploying, and managing container jobs, especially for machine learning tasks. Besides this, it includes some main setup tasks such as performance calibration and controller managing, which in turn establish better-optimized performance of ML apps [24]. Thus, the execution container is utilized for any machine learning job of any size.

### 4.2. AWS ECS Practices and Automation

Deploying machine learning models to the production setting is a difficult operation that involves carefully putting up and running the infrastructure. This issue is fixed by leveraging the application container orchestration managed service of Amazon Web Services (AWS) Amazon Elastic Container Service (ECS), which automates application deployment and scaling. Developers can implement deployment configurations through Infrastructure as Code practices using the provided AWS SDK or CloudFormation tools [20]. Simplification of inner workings of infrastructure while leveraging abstraction, ECS helps to quicken the process of development and the delivery of the Machine Learning apps that are ready to work. In addition, it supports various kinds of deployment structures, such as continuous integration and delivery, and is more suitable in a DevOps setting [23]. Using ECS, the organization can accelerate time to market with automated provisioning/deployment pipelines, which later allows the organization to liquidate the operation of the ML applications by quick deployment.

Figure 5 shows the structure of AWS ECS, including components such as capacity provisioning, controller management, and application lifecycle. In its capacity, it simplifies infrastructure complexity and automates deployment processes, making applications possible to be deployed, managed, and scaled like ML models.

### 4.3. Strengths
*4.3.1. Flexibility with AWS ECS for Docker*

Containerization on AWS Elastic Container Service (ECS) offers organizations a high level of flexibility since containers are highly portable and lightweight, thus making them suitable for bundling applications. Containers enable individual developers to test their applications locally, meeting the same behavior in all environments [17].

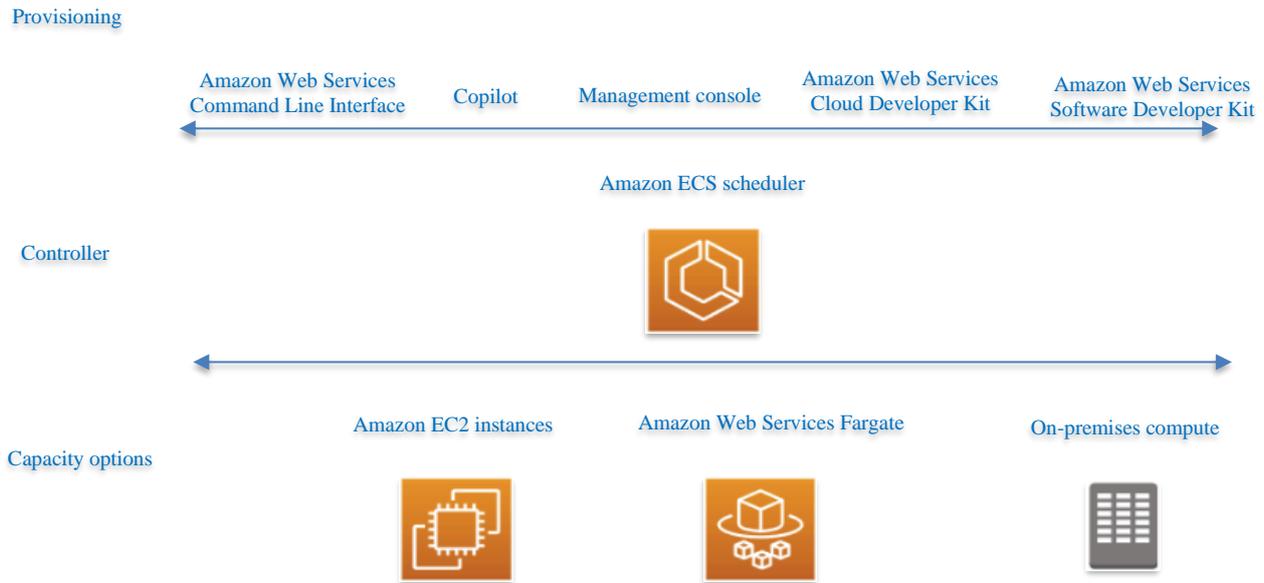

Fig. 5 AWS Elastic container service [34]

Besides, ECS supports Docker containers, allowing them to be used with various tools and frameworks commonly used in containerized applications. This leads to the idea that development staff can use existing containerization workflows and toolchains without extra friction within the AWS environment [18]. ECS supports infrastructure customization through the EC2 launch type or serverless container deployments via the Fargate launch type. It offers a degree of performance variability to fulfill every application requirement and operational preference [22].

*4.3.2. Infrastructure control and horizontal scaling using AWS ECS*

Based on the cloud hosting service, AWS ECS is a thorough and powerful control layer tightly integrated with the infrastructure. Conversely, the users of this application can cater to their environment settings based on personal preferences. ECS, one can set up computing resources, network rules, and security policies [15]. This enables low granular targeting of infrastructure resources for efficiency purposes with cost optimization objectives. Moreover, its ability to scale horizontally allows applications to scale out dynamically or based on demand. Notably, this feature leads to better workload balancing and stable output consumption during users' peak consumption [15]. As such, integration of the infrastructure control systems with horizontal scalability in ECS ensures that capacity management and monitoring of containerized applications are done efficiently.

*4.4. Weaknesses*
*4.4.1. Container Management: A Complication to the Operational Overheads*

These consequences increase the complexity of handling containers en masse and make managing containerized applications inconvenient for organizations. Robust management frameworks must be implemented to effectively control, scale, and monitor containers deployed within an organization [3]. Along with the rise in the number of containerized systems and the complexity of their scale, it becomes imperative to avail automation with an orchestration technique for operational functions such as resource allocation, network configuration, and security control [5]. On top of that, there is a need to provide infrastructure for containerization, which includes infrastructure provisioning, managing container run-time, and networking, which can impact overall system performance and resource utilization [21]. The more important aspects of containerization are for the management to handle operations and possible overheads and implement procedures to guarantee smooth flow when scaling.

*4.4.2. Budgeting, Technological Measures, and the Evaluation of Performance on Containerized Implementation*

Performance comparison of the containers with other strategies like serverless architectures or the mainstream virtual machine-based alternatives is a must when assessing the kind of deployment options based on the containers. The advantages of this approach are, for instance, independence from cloud-based resources and adaptability for different platform architectures together with scalability, but performance will also have to be thought-through before implementation [17]. Some other aspects contributing towards the lowered performance include the lag time it takes for applications to start running following the assignment of IP addresses, increased latency when the sender's request message gets passed along the transmission lines between it and the receiver, and extra cost borne by the service provider during the use period as compared to other AWS services such as Lambda or SageMaker [28]. While performance metrics and cost implications are considered together, more accurate decisions can be made based on the most applicable way of deploying the application to maintain high performance versus low prices.

## 5. Comparative Analysis

### 5.1. Use Case Scenarios and Scalability Scenarios

In the case of managed machine learning platforms, AWS SageMaker is the right choice. It might be utilized to build predictive models for customer segmentation by e-commerce or recommender systems in media streaming platforms [17]. The scalability of a cloud service is its ability to take different kinds of workloads and apply them to scenarios with changing demand. However, AWS's Lambda would be the best option for distributed event architectures that need instant scaling on an event-driven basis [5]. Therefore, it is well used in the applications of IoT, where data is processed right away or in any sudden situation of a large influx of traffic in web applications.

### 5.2. Performance and Cost Evaluation

AWS Lambda meets the needs for quick responsiveness and short-time event-driven tasks, for which fast implementation is required, rendering its performance a factor that stands out. Take, for instance, when discussing low latency scenarios such as request-response interaction type applications [3]. Nevertheless, various highly involving computational tasks can be more suitable for AWS ECS and AWS SageMaker, especially because of their hardware configuration and container-based environment. Compared with any other instances Amazon Web Services provides, its pay-per-use model goes well with sporadic workloads or low-traffic applications. Consequently, it is cost-effective compared with any of them. On the other hand, those organizations that plan on implementing continuous AWS deployments may benefit more from pricing models that are predictable in time [5]. It is based on the structure provided by Amazon Web Services SageMaker and Amazon ECS, which can help optimize costs further.

### 5.3. Use Case-Based Focused Design

The nature of the projects often determines the selection of the exemplary service. In some cases, when simple use of Lambda is necessary alongside sped-up development cycles and reduced costs, they can choose AWS Lambda over the others. In this scenario, developers do not need to manage the infrastructure because of the serverless design, so they can devote time to writing the code and not provisioning the resources [3]. On the contrary, AWS SageMaker or ECS might be versatile enough when it needs better control over the infrastructure configuration settings or support for more complicated ML workflows [3]. Also, the selection of the most suitable deployment service for each use case is affected not only by the matters described above but also by data privacy, regulatory compliance, and integration with existing systems. To state the obvious, a good decision in selecting an AWS model deployment service would involve a thorough examination of the performance, scalability, and cost with a clear focus on the specific requirements of a project.

Table 1.0 Summary of a comparative analysis of AWS model deployment services

| Feature | AWS SageMaker | AWS Lambda | AWS ECS |
|---|---|---|---|
| Deployment Type | Fully managed | Serverless | Container Orchestration |
| Functionality | Build, Train, and Deploy ML Models | Run Code in Response to Events | Deploy, Manage, and Scale Containerized Applications |
| Integration | Seamless integration with other AWS services | Integration with various AWS services | Integration with AWS services like EC2, ECR, and more |
| Scalability | Horizontal and Vertical Scaling | Automatic scaling based on demand | Horizontal Scaling |
| Performance Metrics | Model training time, Inference speed, Resource utilization | Execution time, Cold start latency, Resource consumption | Container startup time, Resource utilization |
| Cost Implications | Pay-as-you-go pricing model | Pay-per-use pricing model | Pay-as-you-go pricing model |
| Strengths | Managed infrastructure, Built-in algorithms and frameworks, Auto-scaling | Cost-effectiveness, Auto-scaling, Integration with the AWS ecosystem | Flexibility, Infrastructure control, Horizontal scaling |
| Weaknesses | Cost implications, Limited customization | Execution time limits, Resource constraints | Operational complexity, Overhead |

## 6. Conclusion

In conclusion, the analysis focused on SageMaker vs. Lambda, and ECS emphasized their specific characteristics and features. SageMaker lets us have a scalable infrastructure at a low price. On the other hand, Lambda is perfect for event-driven, cheaper tasks. In conjunction with this, ECS enables the corresponding scaling up and down of those deployed services based on their demand. To make the right choice, one should consider the following factors: performance, scalability, customization, and cost according to the project's specifications. The future technology of AWS deployment is intended to become more automated and optimized for performance. Future improvements may involve coordinating advanced technologies that will be fast in the implementation processes. Clear goals and requirements are the necessary conditions for companies in terms of which service to opt for in the process of project development.